\documentclass[nofootinbib]{revtex4}
\usepackage[T1]{fontenc}
\usepackage{amsmath,amssymb}
\usepackage{epsfig}
\usepackage{dcolumn}
\usepackage{graphicx}
\usepackage[usenames,dvipsnames]{color}
\usepackage{slashed}
\usepackage[colorlinks,citecolor=blue]{hyperref}
\usepackage{pdfpages}
\usepackage{float}
\usepackage{adjustbox}
\usepackage[autostyle]{csquotes}
\begin{document}
\title{\boldmath Uncertainties in the oscillation parameters measurement due to multi-nucleon effects at NO$\nu$A experiment}
\author{ Paramita Deka$^{1}$ \footnote{E-mail: paramitadeka@gauhati.ac.in}, Jaydip Singh$^{2}$ \footnote{E-mail: jdsingh@fnal.gov}, Neelakshi Sarma$^{1}$, Kalpana Bora$^{1}$ \footnote{kalpana@gauhati.ac.in}} 

\affiliation{Department Of Physics, Gauhati University, Assam, India$^{1}$}
\affiliation{Department Of Physics, Lucknow University, Uttar Pradesh, India$^{2}$}

\begin{abstract}

In this work, we investigate the role of multi-nucleon (MN) effects (mainly 2p-2h and RPA) on the sensitivity measurement of various neutrino oscillation parameters, in the disappearance channel of NO$\nu$A (USA) experiment. Short-range correlations and detector effects have also been included in the analysis. We use the kinematical method of reconstruction of the incoming neutrino energy, both at the near and far detectors. The extrapolation technique has been used to estimate oscillated events at the far detector. The latest global best fit values of various light neutrino oscillation parameters have been used in the analysis. We find that MN effects increase uncertainty in the measurement of neutrino oscillation parameters, while lower detector efficiency is reflected in more uncertainty. This study can give useful insight into precision studies at long-baseline neutrino experiments in future measurements.

\end{abstract}
\maketitle

\section{\label{sec:level1}Introduction}

Precise and adequate knowledge of neutrino scattering cross-sections and nuclear effects in them is very important to reduce the systematic uncertainties in neutrino beam oscillation experiments. The insufficiency in our present understanding of these effects inflicts the precision measurements of yet unknown neutrino oscillation parameters and some other experimentally observed anomalies in the neutrino sector. In current and future neutrino oscillation experiments \cite{Fukuda:1998mi, Fukuda:2002pe, Ahmad:2002jz, Eguchi:2002dm, Michael:2006rx, Abe:2011fz, An:2012eh, Ahn:2012nd, Abe:2017vif, DUNE:2020txw, Ayres:2004js} on neutrino oscillation, nuclear effects in neutrino interactions are one of the principal sources of systematic uncertainties. From the recent result of the T2K \cite{Abe:2018wpn} appearance channel, it is observed that nuclear effects are the largest contributors among all the systematic errors. One of the main contributions to this systematic uncertainty comes from the description of nucleon correlations in the initial state which may induce a 2-particle-2-hole (2p2h) effect in the final state. Due to the presence of such interactions excess events are observed in recent neutrino experiments. The 2p-2h effect is dominated by the meson exchange current (MEC) which involves 2 nucleons, or 2-body current. Nuclear effects (due to the presence of many nucleons in the target) influence both the neutrino scattering cross-section and kinematics of the final state. Due to a poor understanding of nuclear effects in neutrino scattering, uncertainty regarding cross-section models increases. The nuclear effects include Fermi motion of the nucleons, the binding energy of the nucleon in the nucleus, Pauli blocking, and final state interactions (FSIs). FSI further includes re-scattering of the outgoing particles with the nuclear remnant. Nuclear effects (along with FSI) and detector details bias or smear the reconstructed neutrino energy, so a good understanding of these is important.\\

 Long-baseline neutrino oscillation experiments usually use a two detector setup. A small near detector (ND) is placed close to the neutrino production target which constrains neutrino-nucleus interaction cross-sections and neutrino flux and is useful in reducing systematic uncertainties. To observe the neutrino oscillations, a larger far detector (FD) is installed at a larger distance. By comparing the unoscillated and the oscillated flux, the oscillation probability can be measured which later helps to calculate mixing angles, mass squared differences, etc. These experiments aim to measure the six neutrino oscillation parameters-two mass splittings - $\Delta m^{2}_{21}$, $\Delta m^{2}_{32}$, three mixing angles - $\theta_{12}$, $\theta_{23}$, $\theta_{13}$ and one CP phase $\delta_{CP}$, and are designed to measure two channels - appearance and disappearance channels. A neutrino changes its flavor from $\alpha$ to $\beta$ ($\nu_{\alpha}\rightarrow\nu_{\beta}$) during its journey from one detector to another and the appearance experiments measure the probability of appearance $P_{\alpha\rightarrow\beta}(E)$ of this neutrino flavor $\beta$ as a function of energy at a given distance. The disappearance experiment, on the other hand, looks at the probability of disappearance $P_{\alpha\rightarrow\alpha}$(E) of the flavor $\alpha$ at a given distance from the source. Out of these six neutrino oscillation parameters, four parameters $\theta_{12}$, $\theta_{13}$, $\Delta m^{2}_{21}$ and $|\Delta m^{2}_{32}|$ are determined precisely from the experiments, while Octant of the atmospheric mixing angle $\theta_{23}$, leptonic CP-violating phase $\delta_{CP}$ and mass hierarchy remains to be fixed.\\

 Ongoing T2K and NO$\nu$A \cite{Ayres:2004js} experiments are expected to measure these unknown parameters. From the published data of these two experiments in 2018 and 2019 \cite{Abe:2017vif, Abe:2018wpn, Acero:2019ksn, Abe:2019vii}, the T2K best-fit point is at $\sin^{2}\theta_{23}=0.53^{+0.03}_{-0.04}$ for both hierarchies and $\delta_{CP}/\pi=-1.89^{+0.70}_{-0.58}(-1.38^{+0.48}_{-0.54})$ for normal (inverted) hierarchy \cite{Abe:2019vii}. For NO$\nu$A, the best-fit point is at $\sin^{2}\theta_{23}=0.56^{+0.04}_{-0,03}$, and $\delta_{CP}/\pi=0^{+1.3}_{0.4}$ for normal hierarchy. Again, as per the latest results presented at Neutrino 2020 conference, the best-fit point for NO$\nu$A (T2K) is $\sin^{2}\theta_{23}$=0.57 (0.528) and $\delta_{CP}=0.82\pi (-1.6 \pi)$ for normal hierarchy \cite{Miranda:2019ynh}. This tension between the results of two experiments may be an indication of discrepancy among different models.\\

Though sufficient work has been done to understand the role of nucleons in neutrino-scattering still more needs to be done to reduce uncertainties and improve precision. Some of the low energy baseline experiments use a kinematic method of energy reconstruction by assuming a neutron at rest in a quasi-elastic (QE) scattering. Non-QE events may be wrongly identified as QE events due to final state interactions which lead to a wrong value of the reconstructed neutrino energy lower than true incoming energy. From the MiniBooNE and K2K experiments, it came to notice that the cross-section earlier considered as QE contains about 30\% contribution from 2p2h events \cite{Martini:2009uj, Martini:2011wp, Martini:2010ex}. Both charged current (CC) and neutral current (NC) analyses of QE events showed an excess of cross-section than expected from QE scattering. It has been shown that in addition to pion production, any QE scattering event gets background contribution from the presence of multi-nucleon events e.g., 2p2h excitations \cite{Martini:2011wp, Martini:2012uc, Martini:2012fa, Benhar:2013bwa, Gallagher:2011zza, Lalakulich:2012ac} and all other reaction processes \cite{Lalakulich:2012hs} which shifted the events in the lower energy bins. These uncertainties affect the reconstructed energy which further impacts the extraction of oscillation parameters. More details of these effects for MiniBooNE and T2K experiments can be found in Refs. \cite{Lalakulich:2012hs, Meloni:2012fq, Coloma:2013rqa, Coloma:2013tba}. Coloma et al. in \cite{Coloma:2013tba} have shown the impact of different neutrino interactions and nuclear models in neutrino oscillation parameters determination in the $\nu_{\mu}$ disappearance channel. A good agreement of the MiniBooNE data \cite{Martini:2011wp, Martini:2010ex} is obtained by Martini et al. by combining Random Phase Approximation (RPA) with 2p2h contributions. RPA method is a non-perturbative method that is developed to understand the long-range correlation between nucleons inside the nucleus which gives the collective excitation of the nucleus. It is termed as "Random Phase" because the collective excitations of different phases are treated randomly. In GENIE, RPA is included using the Nieves et al. model \cite{Nieves:2004wx} created by the Valencia group to describe the complexity of many-body interactions. This effect modifies the low-lying energy states of the nucleus by changing the potential. Simultaneously it creates a screening effect which in turn reduces the probability for an electromagnetic and weak interaction with those states. Since the probability of weak interaction is less, this results in fewer QE interactions with RPA. Hence it is observed that RPA corrections reduce the cross-sections and it affects largely at lower energies. At higher energies, the RPA reductions become smaller \cite{Nieves:2004wx}, however, RPA suppressions of about 20-30\% for the higher energies can still be found. The RPA suppression also reduces with an increase in lepton effective momentum and it grows with atomic mass number 'A'. From early works on electron scattering data, it has been shown that RPA correction plays a vital role in the peak QE region where these corrections tend to lower the cross-section \cite{Alberico:1986kn}. Similar effects in neutrino-induced interactions are discussed by Kim et al. in Ref. \cite{Kim:1994zea}. In \cite{Nieves:2014lpa}, RPA correlations and 2p2h (multi-nucleon) effects on CC neutrino-nucleus interactions were studied. In a recent publication by MINER$\nu$A collaboration \cite{Rodrigues:2015hik}, they studied the uncertainty in the nuclear models using Carbon target by considering events with and without RPA effects and for RPA+2p2h events and observed that the models require a modification. Further, in more recent results \cite{Filkins:2020xol}, they tuned the 2p2h model which matched with the previous results \cite{Rodrigues:2015hik}. In Ankowski's paper \cite{Ankowski:2015jya}, they compare the kinematic and calorimetric methods of energy reconstruction in the disappearance experiments operating in different energy regimes. Some other similar studies can be found in Refs. \cite{Singh:2019qac, Naaz:2018amr}.\\

Keeping the above issues in mind, we have undertaken this work to investigate the effect of many nucleons in the Carbon target at NO$\nu$A experiment on the extraction of neutrino oscillation parameters. We analyze the disappearance channel for NO$\nu$A experiment in this work. First, we compute the neutrino scattering cross-section for various relevant processes, as explained in detail in later sections. Next, we obtain the event spectrum at ND, followed by that at the FD. Events and migration matrices are computed using a stable version (v-2.12.8) of event generator GENIE \cite{Andreopoulos:2009rq} (Generates Events for Neutrino Interaction Experiments). We have used the kinematic method of energy reconstruction, and Feldman's technique \cite{Feldman:1997qc} to do the $\chi^2$ analysis. Currently many major neutrino baseline experiments such as MINER$\nu$A \cite{Drakoulakos:2004gn}, MINOS \cite{Adamson:2007gu}, MicroBooNE \cite{Chen:2007ae}, NO$\nu$A, T2K and LBNE \cite{Adams:2013qkq} (LBNE is now called DUNE) are using GENIE as event generator. NO$\nu$A uses GENIE (v-2.12.2) \cite{NOvA:2020rbg} tuned to external and NO$\nu$A data for modelling neutrino beam interactions. The detailed procedure to obtain the oscillated event spectrum at FD has been explained in later sections. The unoscillated event spectrum at ND is also used, to take care of the systematics, along with the F/N (Far/Near) ratio of the detectors. Finally, this spectrum of events is used in the sensitivity analysis of extraction of neutrino oscillation parameters. We want to make it clear that we have not attempted to compare NO$\nu$A data with our results - rather, the focus has been to do a simulation to study the effects of multi-nuclear interaction on sensitivity analysis. \\

The novel feature of this work is that we consider the RPA suppression and the multinucleon enhancement at the same time to obtain a complete model to describe the QE interactions. We study the contribution of the full model QE(+RPA)+2p2h MN interactions along with pure QE interaction on FD event distribution, and neutrino and antineutrino sensitivity analysis. This model is then compared to the QE without RPA and without 2p2h. We consider the effect of short-range correlations (SRC) also, which NO$\nu$A ignored in their analyses. We consider two different models for 2p2h interaction - Empirical and Valencia model and also compare the results for them. Detector efficiency and energy resolution are also taken into account, and we found that the oscillation probability curve with 80\% detector efficiency is closest to the curve with no detector effects (than the one with 31.2\% detector efficiency) - it implies that lower efficiency of the detector causes more uncertainty, as it should be. It is seen that the QE(+RPA)+2p2h interactions affect the cross-section significantly and their effect, in turn, is reflected in the event spectrum and extraction of neutrino oscillation parameters alike. The addition of 2p2h to QE produces QE-like events which increase the cross-section and the event distribution, which in turn reduces the contour areas in the sensitivity analysis.\\

The paper has been organised as follows. In section II, we briefly review the necessary details of the NO$\nu$A experiment, and of the multi-nucleon effects (2p2h, RPA) in section III, for the sake of completeness of this work. Section IV contains the necessary physics and simulation details, where we discuss the numerical method of our analysis, detector effects, the extrapolation technique used to obtain the far detector energy spectrum of events, and sensitivity in the extraction of the oscillation parameters. In Section V, we present our results and a discussion on them. This analysis is done for both neutrino and antineutrino incoming beams for different processes: QE (without RPA suppression and 2p2h), and QE(+RPA)+2p2h with two different models for 2p2h contribution. We summarise and conclude the work in Section VI.

\begin{figure}[H]
 \centering\includegraphics[width=7cm, height=5cm]{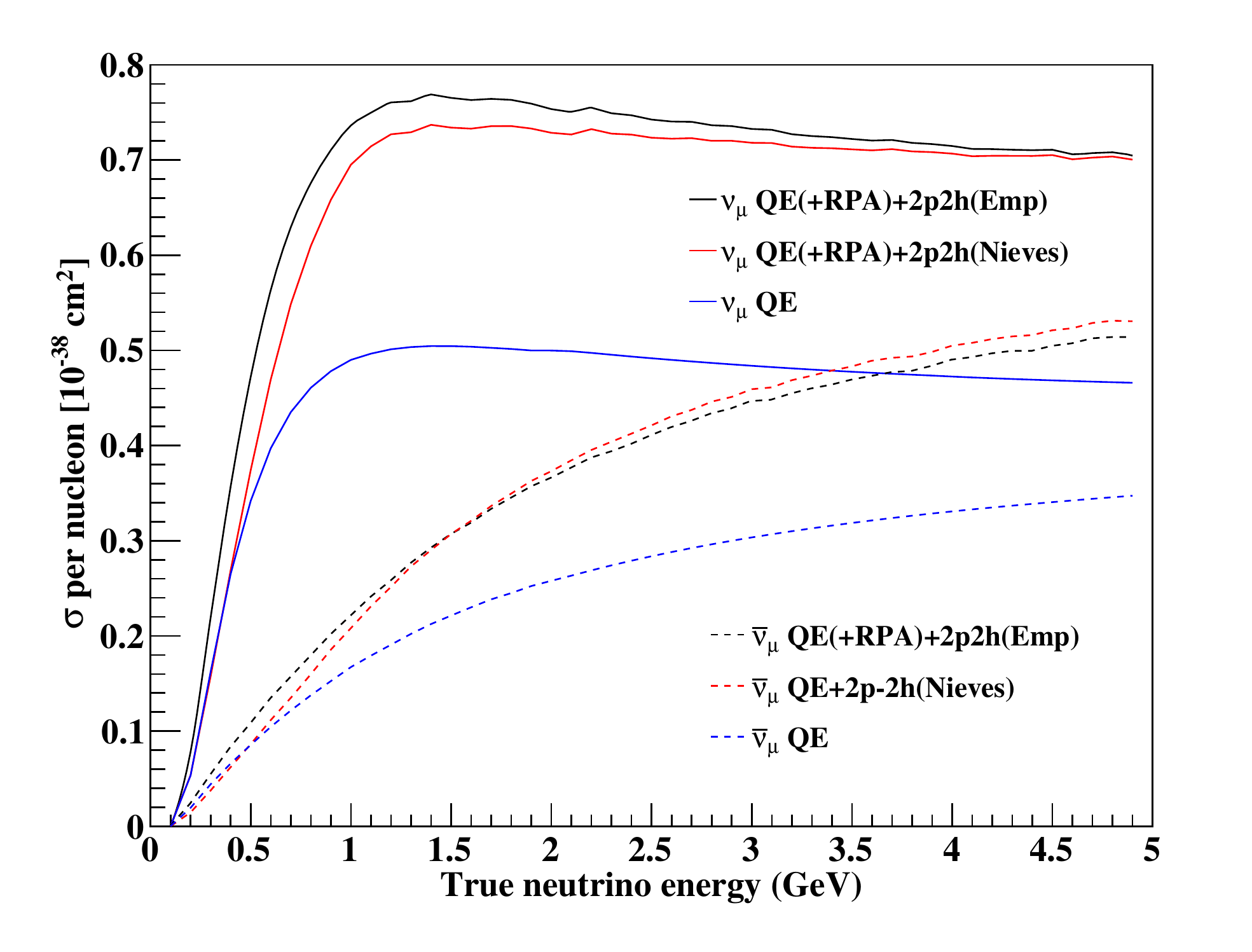}
 \caption{Cross section on Carbon for both neutrino (solid line) and antineutrino (dotted line) as a function of true neutrino energy for QE(+RPA)+2p2h (Empirical, black line), QE(+RPA)+2p2h (Nieves et al., red line), and QE without RPA and 2p2h (blue line) interactions.}
\end{figure} 

\section{The NO$\nu$A Experiment}
\label{sec:1}
The NuMI Off-Axis $\nu_{e}$ Appearance (NO$\nu$A) is a long-baseline neutrino experiment which consists of two functionally identical, segmented, tracking calorimetric detectors - ND and FD to measure the muon neutrino ($\nu_{\mu}$) disappearance probability P($\nu_{\mu}(\bar\nu_{\mu})\rightarrow\nu_{\mu}(\bar\nu_{\mu})$) and electron neutrino ($\nu_{e}$) appearance probability P($\nu_{\mu}\rightarrow\nu_{e}$). The 3.8 m$\times$3.8 m$\times$12.8 m ND has a mass of about 0.3 kilotons and is placed 105 m underground, 1 km from Neutrinos at the Main Injector (NuMI) beam at Fermilab \cite{Acero:2019ksn}, USA. The 15 m$\times$15 m$\times$60 m 14 kiloton FD is located near Ash River, Minnesota at a distance of 810 km from the NuMI \cite{Adamson:2015dkw} target. Both the detectors are placed 14.6 mrad off-axis from the center of the NuMI beam to provide a narrow beam neutrino flux peaked at around $\sim$ 2 GeV and to enhance the sensitivity to $\nu_{\mu}$ disappearance and $\nu_{e}$ appearance processes. A high-intensity neutrino beam is generated by the collision of 120 GeV protons coming from NuMI onto a 1.2 m graphite target. Pions and kaons produced in the target are focused on by two magnetic horns. In the NuMI beam, most of the neutrinos result from the process $\pi^{\pm}\rightarrow\mu^{\pm}+\nu$. The main objective of these two functionally equivalent detectors placing at a long distance is to measure neutrino oscillation, reduce systematic uncertainty, and from these oscillation measurements NO$\nu$A is expected to determine mass hierarchy, leptonic CP-violating phase $\delta_{CP}$, and also resolve the Octant degeneracy of the atmospheric mixing angle $\theta_{23}$ \cite{Bora:2014zwa, Bora:2014kva, Bora:2016tmb}. 

\begin{figure}[H] 
\centering\includegraphics[width=13.5cm, height=6cm]{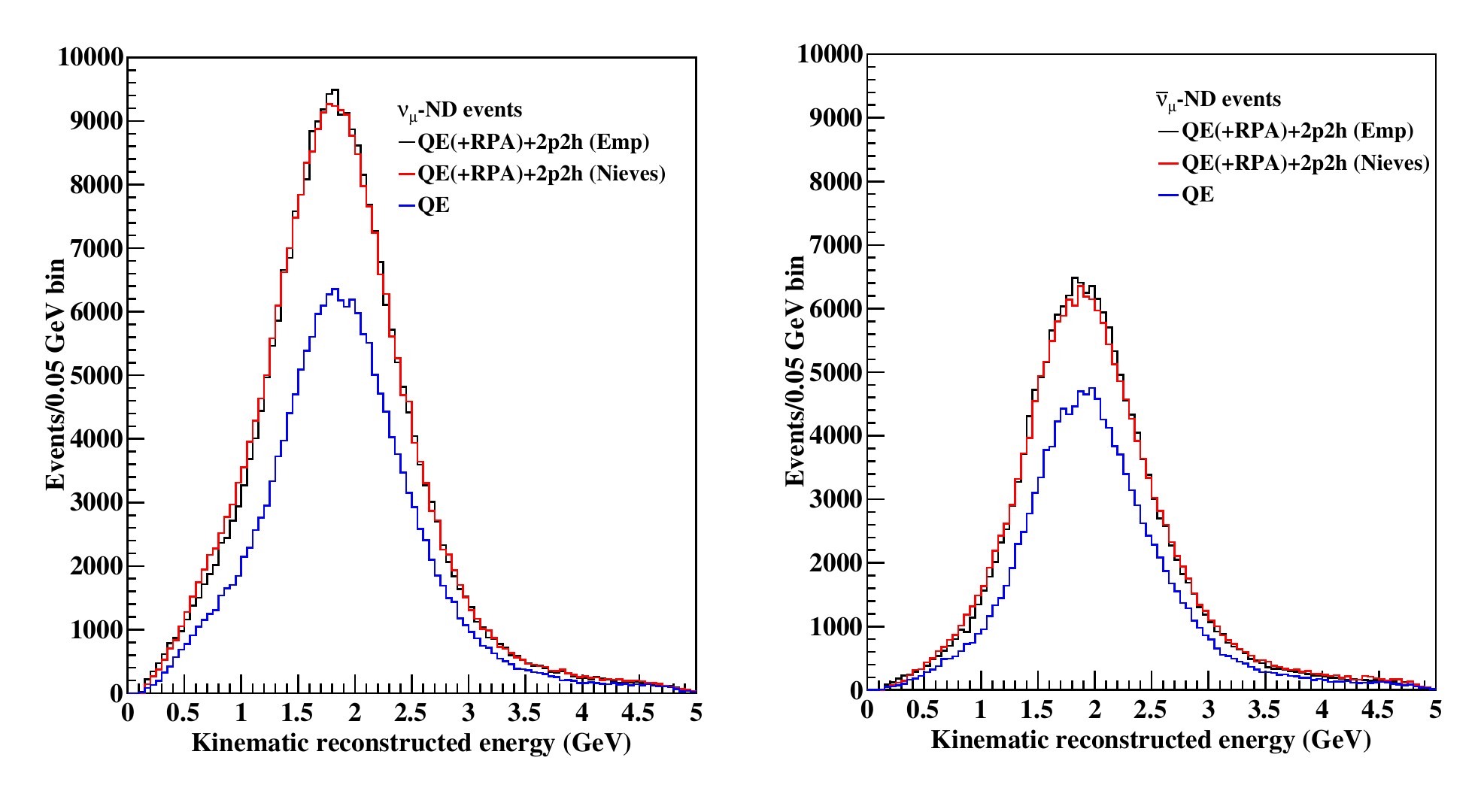}
\caption{Events as a function of reconstructed $\nu_{\mu}$ and $\bar\nu_{\mu}$ for both neutrinos (solid line) and antineutrinos (dotted line) at ND, for three different cases: QE(+RPA)+2p2h (Empirical, black line), QE(+RPA)+2p2h (Nieves et al., red line), and QE without RPA and 2p2h (blue line).}
\end{figure}

\section {Multi-nucleon Interactions}
\label{sec:2}
Currently, the neutrino scattering community is more interested in a process beyond QE and resonance processes, when neutrino scatters off a nucleus. In this process, when a neutrino interacts with the nucleons inside the nucleus exchanging a W boson, it is absorbed by nucleons which results in the knock out of two-particle and two-hole pairs (2p2h) through an exchange of a meson. QE interaction is referred to as 1p1h interaction because one particle is ejected out from the nucleus leaving one empty unoccupied (hole) energy state. Together multi-nucleon excitation and charged lepton in the final state (without pion absorption) are known as quasielastic (QE)-like events. 2p2h events mostly occur in the energy region between QE and resonance production. This process contributes significantly to the neutrino-nucleus scattering and its contribution can be observed in the cross-section plot in Fig. 1. The QE cross-section thus contains a certain proportion of 2p2h and 3p3h excitations \cite{Martini:2009uj, Martini:2011wp, Martini:2010ex, Martini:2012uc}. The nucleon-nucleon correlations in the initial state are often known as short-range correlations \cite{Gallagher:2011zza}. Some works on np-nh contribution to neutrino-nucleus cross-section can be found in - (phenomenological approaches) by Mosel et. al. \cite{Lalakulich:2012hs, Lalakulich:2012ac, Mosel:2014lja}, Bodek et. al. \cite{Bodek:2011ps}. In GENIE two models are available for the MEC process and NO$\nu$A uses both of these - the Empirical and Valencia MEC models \cite{NOvA:2020rbg}. It is clear from their analysis that any MEC model present in GENIE requires significant tuning to reproduce their data. We have used both these models in our analysis for comparison and a better understanding of these issues. \\

In Fig. 1, we have shown how the MN interactions QE(+RPA)+2p2h affect the neutrino-nucleus cross-section for the Carbon target (per nucleon). We have plotted for QE(+RPA)+2p2h (Empirical, black line), QE(+RPA)+2p2h (Nieves et al., red line), and QE without RPA and 2p2h (blue line) for neutrinos (solid lines) and antineutrinos (dotted lines). To see how RPA+2p2h effects along with QE interactions influence the event distributions, we show the spectrum of QE(+RPA)+2p2h and QE events without 2p2h and RPA corrections in ND, both for neutrino and antineutrino in Fig. 2. From Fig. 1, it is evident that due to the presence of MN interactions, the QE(+RPA)+2p2h cross-section is more than the pure QE for both neutrino and antineutrino. This is as expected - because these additional interactions add to the strength (through additional contributions) of the pure QE process, and hence enhance the cross-section. From the comparison of the Empirical and Nieves et al. model for 2p2h, we observe that in the case of the neutrino, QE(+RPA)+2p2h cross-section for Empirical model is more than QE(+RPA)+2p2h cross-section for Nieves et al. model. However, for antineutrino, QE(+RPA)+2p2h (Empirical) cross-section is slightly more than the QE(+RPA)+2p2h (Nieves et al.) cross-section in the lower energy range $\leqslant$ 1.3 GeV. Both the cross-section overlap around 1.5 GeV and after that QE(+RPA)+2p2h (Nieves et al.) cross-section is greater than QE(+RPA)+2p2h (Empirical) in the higher energy range. This difference may be attributed to the nature of physical processes in the two models. There is suppression in the QE cross-section due to RPA corrections in both neutrino and antineutrino scattering. This is also as expected because it has been argued that RPA quenching of cross-sections at low $Q^{2}$ arises due to the repulsive nature of particle-hole forces, and they also cause a shift of cross-section towards larger angles or larger kinetic energy of muons \cite{Martini:2011wp}. As a result of which, a similar feature is observed in Fig. 2 for the event spectrum at the ND. From the left panel of Fig. 2, we observe the event distributions at ND for Empirical and Nieves et al. model for 2p2h interactions in QE(+RPA)+2p2h are almost similar for neutrino in lower and higher energy sides but there is a slight difference in peak position in neutrino and antineutrino case. The ND event distribution for QE(+RPA)+2p2h (Empirical) is slightly more than QE(+RPA)+2p2h (Nieves et al.) in the peak position at around 2 GeV.

\begin{figure}[H] 
 \centering\includegraphics[width=13cm, height=10cm]{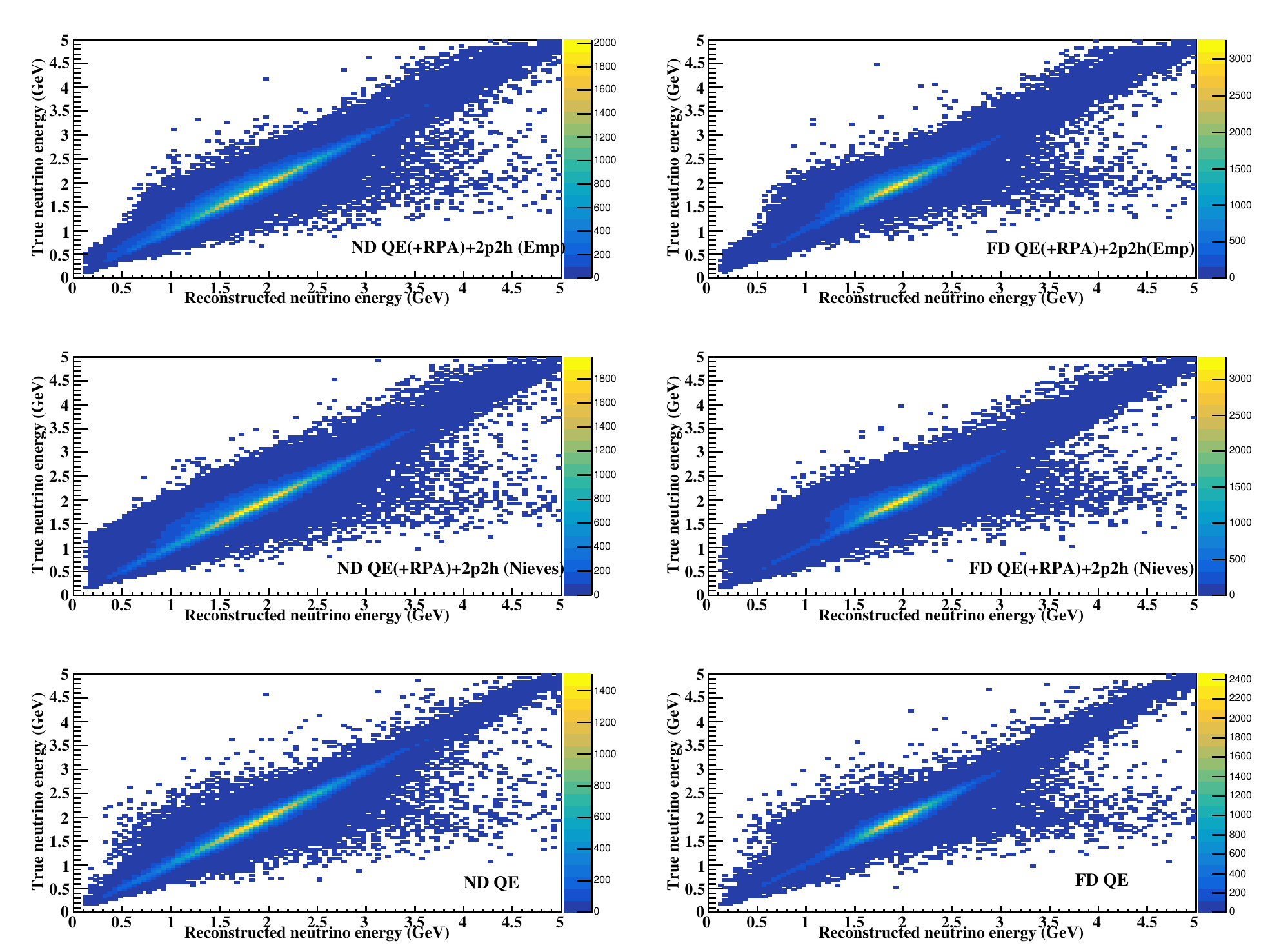}
\caption{Reco-to-true and true-to-reco migration matrices for ND (left) and FD (right) events for three different interactions with FD detector efficiency 80\%. For each detector a 2D histogram of reconstructed energy vs. true energy is created. Using these reco-true matrices, ND neutrino spectrums in bins of reconstructed energy are converted to bins of true energy and vice versa for FD energy spectrum.} 
\end{figure}

\section{Physics and Simulation Details of this work}
\label{sec:3}

For investigating the effect of multi-nucleon interactions on sensitivity measurement of neutrino oscillation parameters, as stated earlier, in this work we consider CC QE, RES, DIS, and 2p2h interactions. We have generated 1 million MC events including these four processes for both ND and FD using NO$\nu$A flux in the energy range 0.0-5.0 GeV to measure the energy spectrum of $\nu_{\mu}$ ($\bar\nu_{\mu}$) CC interactions, and calculated reconstruction energy using kinematic method (equation (2)). From that event sample, we have extracted the desired events. We have used GENIE default models for simulation - QE scattering predicted according to Llewellyn-Smith \cite{LlewellynSmith:1971uhs}, resonance (RES) processes according to Rein-Sehgal model \cite{Rein:1980wg} with 16 resonance models while DIS interactions formulated by Bodek and Yang model \cite{Bodek:2002ps}. For modelling 2p2h interaction, we have used two models-Empirical model \cite{Katori:2013eoa} and Nieves et al. model (Valencia model) \cite{Nieves:2011yp}. Furthermore, we have tuned GENIE's description to include long-range nucleon correlations - the RPA effect with the Nieves et al. model and to include short range correlations we use RFG model modified by A. Bodek and J.L Ritchie \cite{Bodek:1981wr}. For 2p2h Nieves et al. model, we consider the Local Fermi Gas (LFG) to model the initial nuclear state. The electromagnetic form factor can be parameterized by vector form factor BBBA2005 model \cite{Bradford:2006yz} with a variable value of axial mass $M_{A}$ between 0.99-1.2 GeV/$c^{2}$. We have used axial mass $M_{A}$=1.04 GeV/$c^{2}$ \cite{NOvA:2020rbg}. FSI is modeled with the INTRANUKE \cite{Dytman:2009zz} cascade model. RFG is currently used as the main model in the analysis of long-baseline experiments to get high-precision measurements of the neutrino oscillation parameters. Inside the nucleus, the motion of the nucleons can be understood in two different ways in Llewellyn-Smith model and Nieves et al. model. In Llewellyn-Smith model, Global Fermi Gas (GFG) is used to model the nucleus, where the nucleus is assumed to have constant nuclear density and with a uniform momentum distribution in 3-D momentum space. On the other hand, the Nieves et al. model considers the nucleus as a LFG which is a minimal extension of the RFG model. In the LFG model, the Fermi momentum and binding energy depend on the distance, i.e., the radial position of the nucleons in the nucleus. NO$\nu$A uses models developed by the Valencia group for QE and 2p2h interactions and used the LFG model to describe the initial nuclear state in both Valencia QE and 2p2h models. They have simulated RES processes using the Berger-Sehgal model and DIS by Bodek-Yang model \cite{NOvA:2021nfi}. 

\subsection{Oscillation Analysis}
\label{sec:A}

In the neutrino oscillation analysis, the significant background contribution comes from neutral current events which are wrongly identified as CC events. The cosmic background contributes 4.1\% of the selected FD $\nu_{\mu}$ CC events and NC background events contributes 6\% of FD $\nu_{\mu}$ CC events \cite{Adamson:2016xxw}. In general, this contribution is considered to be very low, therefore in our analysis, we have neglected it. The relevant oscillation probability for NO$\nu$A for the disappearance channel can be expressed as \cite{MendezMendez:2019vyt}
\begin{equation}
P(\nu_{\mu}\rightarrow\nu_{\mu})=1-\sin^{2}2\theta_{23}\sin^{2}\Delta_{32}+4\sin^{2}\theta_{23}\sin^{2}\theta_{13}\cos^{2}2\theta_{23}\sin^{2}\Delta_{32}
\end{equation}
where $\Delta_{ij}=\frac{\Delta m^{2}_{ij}L}{4E_{\nu}}$, $\Delta m^{2}_{ij}$ are the mass differences and $\theta_{ij}$ are the mixing angles. 
The true (global best fit) values of the oscillation parameters used in the work are shown in Table I.

\begin{table}[h]
\begin{center}
\begin{tabular}{|c|c|c|}
\hline
parameter & best fit & 3$\sigma$ range \\
\hline
$ \Delta m_{21}^2[10^{-5} eV^{2}]$ & $7.50$ & $6.94-8.14$\\
$ |\Delta m_{31}^2|[10^{-3} eV^{2}]$(NH) & $2.56$ & $2.46-2.65$ \\
$ |\Delta m_{31}^2|[10^{-3} eV^{2}]$(IH) & $2.46$ & $2.37-2.55$ \\
$ \sin^2\theta_{23}/10^{-1}$(NH) & $ 5.66$ & $4.46-6.09$ \\
$ \sin^2\theta_{23}/10^{-1}$(IH) & $ 5.66$ & $4.41-6.09$ \\
$ \sin^2\theta_{13}/10^{-1}$(NH) & $ 2.225$ & $2.015-2.417$ \\
$ \sin^2\theta_{13}/10^{-1}$(IH) & $ 2.250$ & $2.039-2.441$ \\
\hline
\end{tabular}
\end{center}
\caption{3$\sigma$ values of neutrino oscillation parameters taken from \cite{deSalas:2020pgw}.}
\label{tab:data1}
\end{table}

\subsection{Energy Reconstruction}
\label{sec:B}
The neutrino energy is constructed using the kinematic reconstruction method which relies on the kinematics of outgoing lepton $\textit{l}$ (energy and angle). This method of neutrino energy reconstruction is based on the assumptions that the beam particle interacts with a single neutron at rest with constant binding energy and that no other nucleons are knocked out from the nucleus. For a QE event, the neutrino energy can be reconstructed as \cite{Coloma:2013tba}:  

\begin{equation}
E^{QE}_{\nu}=\frac{ 2(M_{n}-E_{b})E_{\textit{l}}-(E^{2}_{b}-2M_{n}E_{b}+\Delta M^{2})}{2(M_{n}-E_{b}-E_{\textit{l}}+\textit{p}_{\textit{l}}\cos\theta_{\textit{l}})}
\end{equation}\\

where $\Delta M^{2}=M^{2}_{n}-M^{2}_{p}+\textit{m}^{2}_{\textit{l}}$ and $E_{b}$ is the average binding energy of the nucleon inside the nucleus. It may be noted that there are examples of using the same formula (Eq.(2)) for both neutrino and antineutrino energy reconstruction, with same/slightly different values of binding energy. For example, in \cite{Ankowski:2015jya}, Ankowski et. al. used the same formula and binding energy value for both neutrino and antineutrino, while in \cite{Wilkinson:2016wmz}, they used two slightly different values for neutrino and antineutrino binding energy. In this work, we use $E_{b}$=25 MeV for carbon for both neutrino and antineutrino \cite{MINERvA:2019ope}. $M_{n}$ is the free neutron rest mass, $E_{\textit{l}}$ and $\theta_{\textit{l}}$ are the energy and angle of the outgoing lepton respectively. This equation is exact only for QE interaction with the neutron at rest.

\subsection{Extrapolation Technique}
\label{sec:C}

After selecting the ND events, the next step is to predict the corresponding number of interactions in the FD. The technique of predicting the FD energy spectrum using ND data is known as extrapolation. NO$\nu$A uses Monte Carlo simulation to extrapolate the neutrino energy spectrum measured in the ND to the FD. Monte Carlo simulations for ND and FD are generated separately under the assumption of no neutrino oscillation. The $\nu_{\mu}$($\bar\nu_{\mu}$) signal spectra at the FD are predicted for both neutrino and antineutrino beams separately based on the constrained $\nu_{\mu}$ ($\bar\nu_{\mu}$) event predictions in the ND. The discrepancy between ND and FD spectra mainly is due to the geometric effects. Using this technique, the smearing effect of imperfect energy resolution in both the detectors can be detected. This is done by creating a 2D histogram between the reconstructed energy and true energy in ND and FD detectors. Further, these histograms are used to re-weight the reconstructed and true spectrum to get the extrapolated FD events. The extrapolation of $\nu_{\mu}$ ($\bar\nu_{\mu}$) CC signal is done in bins of true energy. The full procedure of extrapolation can be explained as follows \cite{NOvA:2016vij, NOvA:2017ohq, Pershey:2018gtf}:

\begin{enumerate} 
 \item The reconstructed ND energy spectrum (X-axis) is converted to the spectrum in bins of true of energy (Y-axis) with the help of a reco-to-true migration matrix from simulation. Each element of the matrix represents the probability that a neutrino with reconstructed energy $E_{i}$ came from a neutrino with true energy $E_{j}$ where $j$ is the index over true energy, i.e., for each given value of the reconstructed energy, a probability distribution in true energies can be computed. As the number of events before and after migration should remain the same, it is required that sum of probability densities for given true energy be normalized to the unity. The migration matrices used for three different processes in ND and FD are illustrated in Fig. 3.  
 
 \item The geometric effect (e.g., angular acceptance, decay kinematics, beamline geometry, focusing of the particles) is taken care of by multiplying this true energy spectrum with a far-to-near ratio to get FD unoscillated prediction in bins of true energy. The main advantage of this approach is the reduction of many systematic uncertainties that affect both detectors, resulting in a smaller error on the oscillation measurements. This ratio also includes the efficiency and acceptance differences between the two detectors, differences in detector fiducial volumes, and differences in the flux at the ND and FD \cite{Pershey:2018gtf}.
  
  \item Then this true energy spectrum is multiplied with oscillation probability to re-weight the spectrum.
 \item Finally, using underlying energy distributions from simulated neutrino interactions in the FD (true-to-reco migration matrix) this oscillated spectrum in bins of true energy is mapped to the spectrum in bins of reconstructed energy. 
\end{enumerate}

The extrapolated FD energy spectrum predicted from the simulated ND energy spectrum is shown in Figs. 4 and 5 for both neutrino and antineutrino and both the mass hierarchies. Also, the features seen in Fig. 1 and Fig. 2 are reflected in the FD events spectrum in Fig. 4 and Fig. 5 too. The MN interactions are seen to affect the FD events spectrum quite significantly, they change the amplitude as well as the position of the peaks in the FD events spectrum. We know that amplitude and phase in oscillation probability are reflected in measurements of oscillation parameters ($\Delta m^{2}$, $\theta$ respectively) in the experiments. Since QE(+RPA)+2p2h interaction increases the amplitude, it is expected to decrease the uncertainty in the measurement of oscillation parameters. From the corresponding spectrum of events at FD, after using the extrapolation technique, in Fig. 4, the oscillatory nature of neutrino oscillation probability can be seen. It may be noted that the results shown in Figs. 4 and 5 are the outcomes of folding of flux, cross-section (Fig. 1), ND events spectrum (Fig. 2), and oscillation probability as we know neutrino event rates are a convolution of neutrino flux, neutrino-nucleus cross-section, nuclear effects, and detector response. It can be seen that the MN effects change the cross-section for neutrino more, than for antineutrino, and this feature has been noted earlier also \cite{Martini:2010ex}, where they attributed this effect to possible cancellation of some MN terms in antineutrino case. In the second peak position near 1 GeV of the FD spectrum for neutrino, QE(+RPA)+2p2h (Nieves et al.) events are slightly more than QE(+RPA)+2p2h (Empirical) and shifted towards lower energy side while in the third peak position at around 2.2 GeV, the event distribution for QE(+RPA)+2p2h (Empirical) is more than QE(+RPA)+2p2h (Nieves et al.). Similar results are observed in case of antineutrino also. Both for neutrino and antineutrino, there is a clear distinction among the three curves - pure CC QE, and QE(+RPA)+2p2h (Empirical) and QE(+RPA)+2p2h (Nieves et al.) interactions, in medium energy ranges, say $0.5 \leqslant E_{\nu}\leqslant3$ GeV and $0.5\leqslant E_{\nu}\leqslant 4.5$ GeV respectively. It implies that at energies beyond these ranges, the effect of these interactions diminishes. These observations can be attributed to the fundamental structure and nature of these interactions in these energy ranges as also discussed in the introduction. RPA is a long-range effect, which becomes significant at small $Q^{2}$ only, and np-nh forces also cannot be significant for the whole range of neutrino energies.

 \begin{figure}[H] 
\centering\includegraphics[width=13.5cm, height=5.5cm]{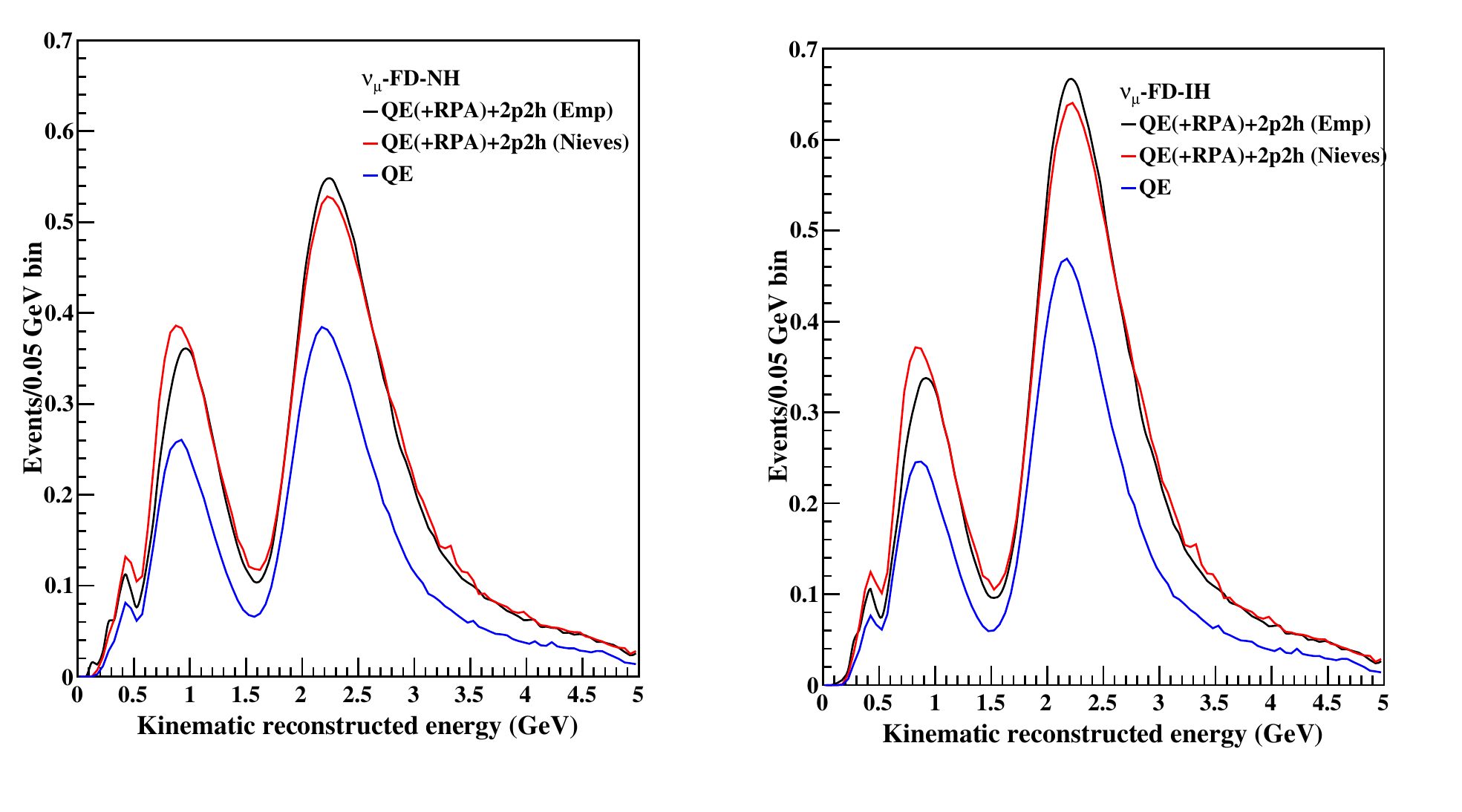}
\caption{Left and right panels show respectively, extrapolated FD events for neutrino, as a function of reconstructed $\nu_{\mu}$ energy for NH and IH. Events are shown for three different cases: QE(+RPA)+2p2h (Empirical, black line), QE(+RPA)+2p2h (Nieves et al., red line), QE without RPA and 2p2h(blue line).} 
\end{figure}

\begin{figure}[H] 
\centering\includegraphics[width=13.5cm, height=5.5cm]{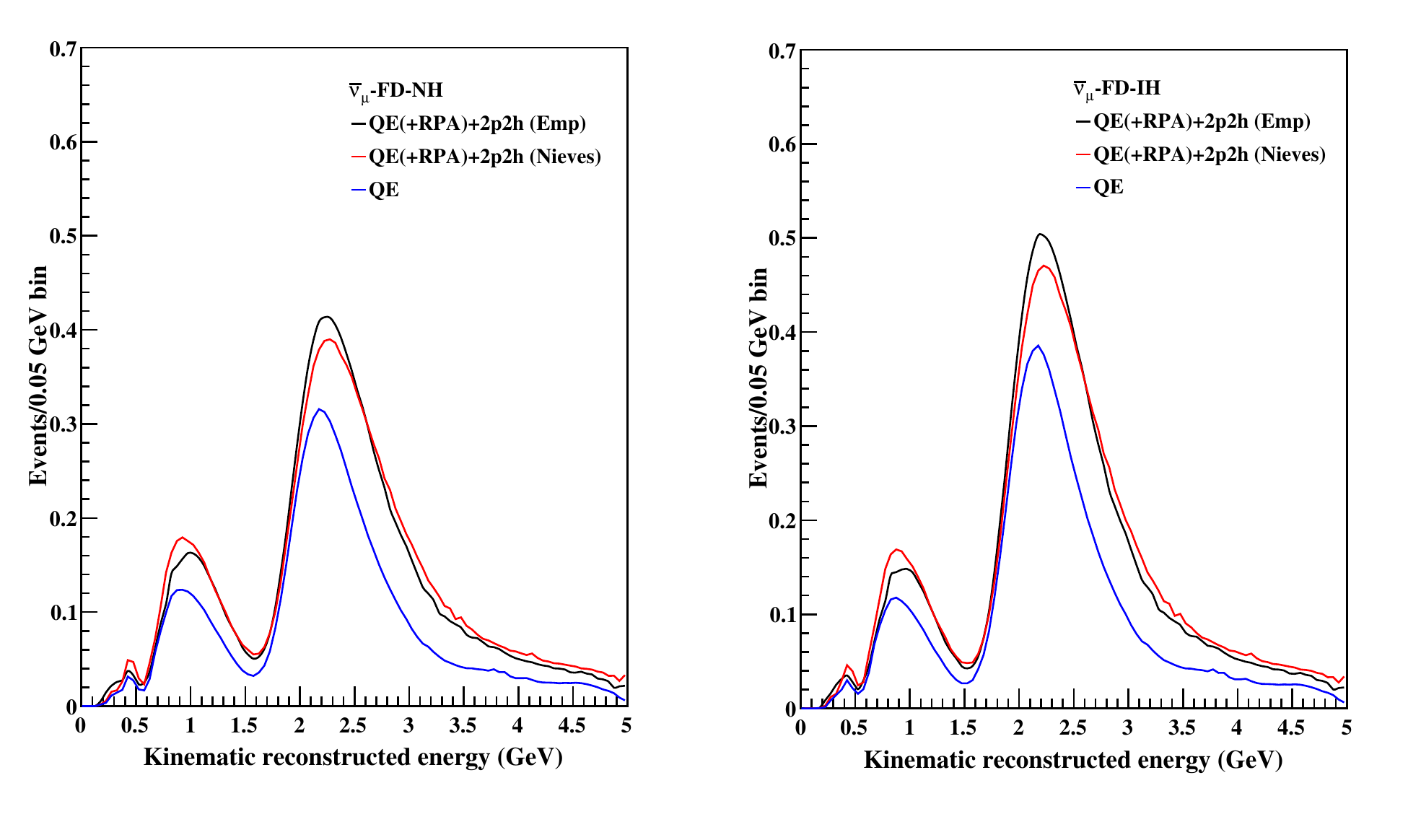}
\caption{Left and right panels show respectively, extrapolated FD events for antineutrino, as a function of reconstructed $\bar\nu_{\mu}$ energy for  NH and IH. Events are shown for three different cases: QE(+RPA)+2p2h (Empirical, black line), QE(+RPA)+2p2h (Nieves et al., red line), QE without RPA and 2p2h(blue line).} 
\end{figure}
 
\subsection{Detector Effect}
\label{sec:D}

The neutrino energy resolution at NO$\nu$A experiment at the FD is 9.1\% (8.1\%) for $\nu_{\mu}$ CC ($\bar\nu_{\mu}$ CC) events. The efficiency of selection of $\nu_{\mu}$ ($\bar\nu_{\mu}$) events is 31.2\% (33.9\%) relative to true interactions in the fiducial volume which results in 98.6\% (98.8\%) purity at the FD during neutrino (antineutrino) beam \cite{Acero:2019ksn}. For the disappearance channel, both $\nu_{\mu}$ and $\bar\nu_{\mu}$ are considered as the signal. We have considered detector efficiency 31.2\% ($\nu_{\mu}$) and 33.9\% ($\bar\nu_{\mu} $) for sensitivity analysis and also compare it with the other two values for comparison - just to see what happens if the detector performance is improved by 50\% or 80\%.  

\subsection{Sensitivity analysis of neutrino oscillation parameters}
\label{sec:E}
Next, we focus on the sensitivity study of the neutrino oscillation parameters $\theta_{23}$ and $\Delta m^{2}_{32}$, applying the result of the above-discussed analysis of multi-nucleon effects (as the $\nu_{\mu}$ disappearance channel is sensitive to $\sin^{2}\theta_{23}$ and $|\Delta m^{2}_{32}|$). This would bring us closer to the aim of this investigation - how the multi-nucleon interactions in the target nucleus affect the measurement of neutrino oscillation parameters at NO$\nu$A experiment. The allowed confidence level regions in parameter spaces are obtained using the Feldman-Cousins method \cite{Feldman:1997qc}. We calculate the acceptance region of each point in the $\sin^{2}\theta_{23}$-$\Delta m^{2}_{32}$ plane by performing a Monte Carlo simulation of the results obtained from a large number of simulated experiments for the given set of unknown physical parameters with the known neutrino flux of the actual experiment using,

\begin{equation}
\Delta \chi^{2}=2\displaystyle\sum_{i}\Big[\mu_{i}-{\mu_{best}}_{i}+{\mu_{best}}_{i}ln\Big(\frac{{\mu_{best}}_{i}+b_{i}}{\mu_{i}+b_{i}}\Big)\Big]
\end{equation}

\begin{figure}[H] 
 \centering\includegraphics[width=17cm, height=12cm]{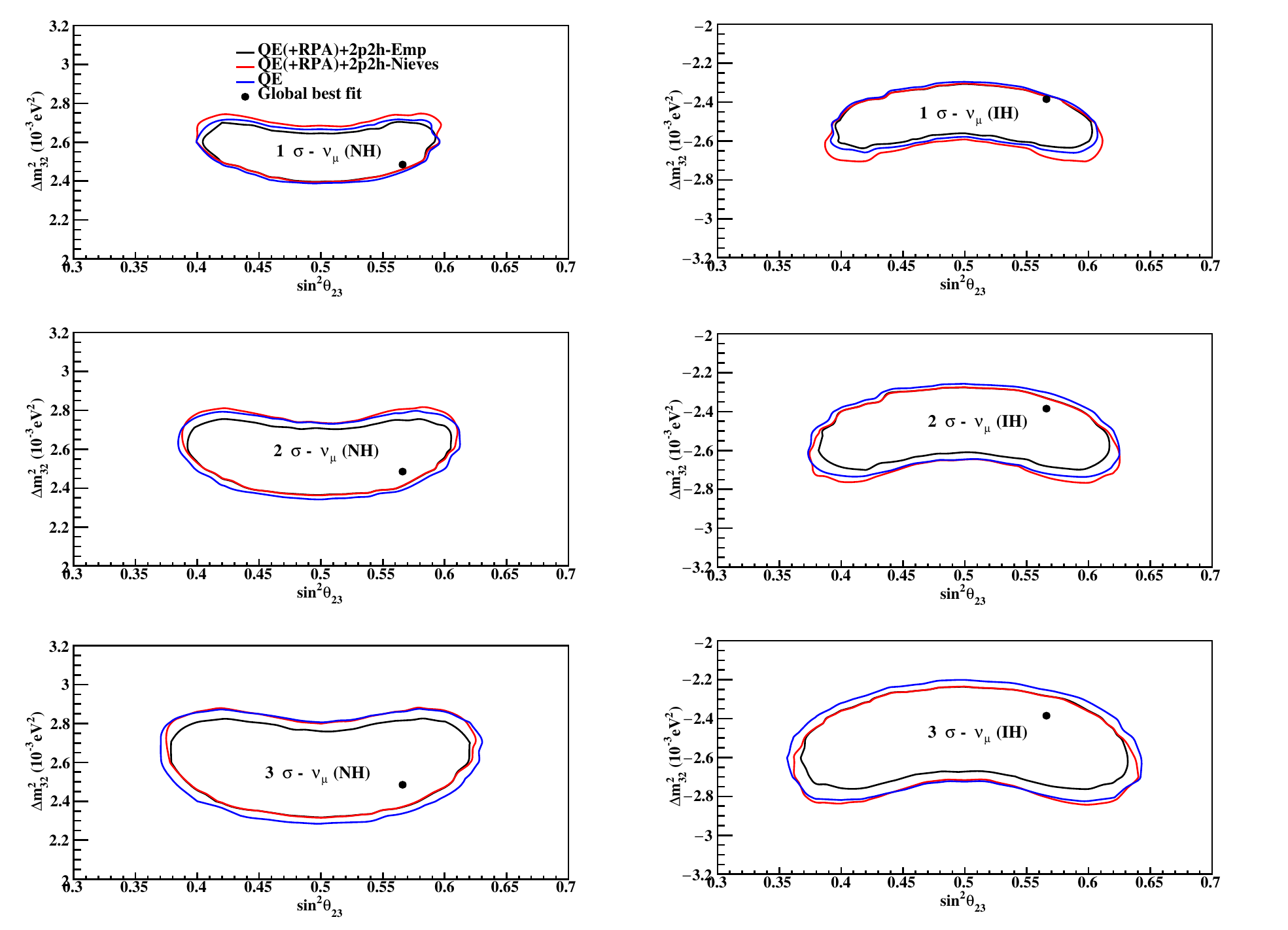}
 \caption{Comparison of 1$\sigma$, 2$\sigma$ and 3$\sigma$ contours in $\Delta m^{2}_{32}$ vs $\sin^{2}\theta_{23}$ plane for three different cases QE(+RPA)+2p2h (Empirical, black line), QE(+RPA)+2p2h (Nieves et al., red line), QE without RPA and 2p2h (blue line) for NH (left panel) and IH (right panel) for neutrino without detector response. The global best-fit point is shown by a black marker.} 
\end{figure}

where the sums runs over all bins, $b_{i}$ is the mean expected background, $\mu_{i}$ is the number of predicted events for
 $i^{th}$ bin, and ${\mu_{best}}_{i}$ is the number of observed events for $i^{th}$ bin in FD data. The predicted $\mu_{i}$ depends on the values chosen for $\theta_{13}$, $\theta_{23}$, $\Delta m^{2}_{32}$, $\Delta m^{2}_{21}$. We have neglected the background contribution. We determine the value of $\Delta\chi^{2}$ for each simulated experiment and $\Delta\chi^{2}_{c}(\sin^{2}\theta_{23}, \Delta m^{2})$ is the critical value corresponding to desired precision. After the data analysis, comparing $\Delta\chi^{2}(N|\sin^{2}\theta_{23}, \Delta m^{2}_{32})$ with $\Delta\chi^{2}_{c}$, the acceptance region for all point is obtained by imposing the condition \par

\begin{equation}
\Delta\chi^{2}(N|\sin^{2}\theta_{23}, \Delta m^{2}_{32})< \Delta\chi^{2}_{c}(\sin^{2}\theta_{23}, \Delta m^{2}_{32})
\end{equation}
The log-likelihood function is calculated for different oscillation parameters and 2D surface of $\chi^{2}$ vs $\Delta m^{2}_{32}$ and $\sin^{2}\theta_{23}$ is formed. The values of $\Delta\chi^{2}$ for different $\sigma$ and confidence intervals in 2D case is shown in Table 2. 

\begin{table}[h]
\begin{center}
\begin{tabular}{|c|c|c|}
\hline
Sigma & CL (\%)) & 2D \\
\hline
1$\sigma$ & $68.3$ & $2.30$ \\
2$ \sigma$ & $95.45$ & $6.18$ \\
3$ \sigma$ & $ 99.73$ & $11.83$ \\
\hline
\end{tabular}
\end{center}
\caption{$\Delta\chi^{2}$ for two dimensions for different $\sigma$ and confidence levels.}
\label{tab:data1}
\end{table}

\begin{figure}[H] 
 \centering\includegraphics[width=17cm, height=12cm]{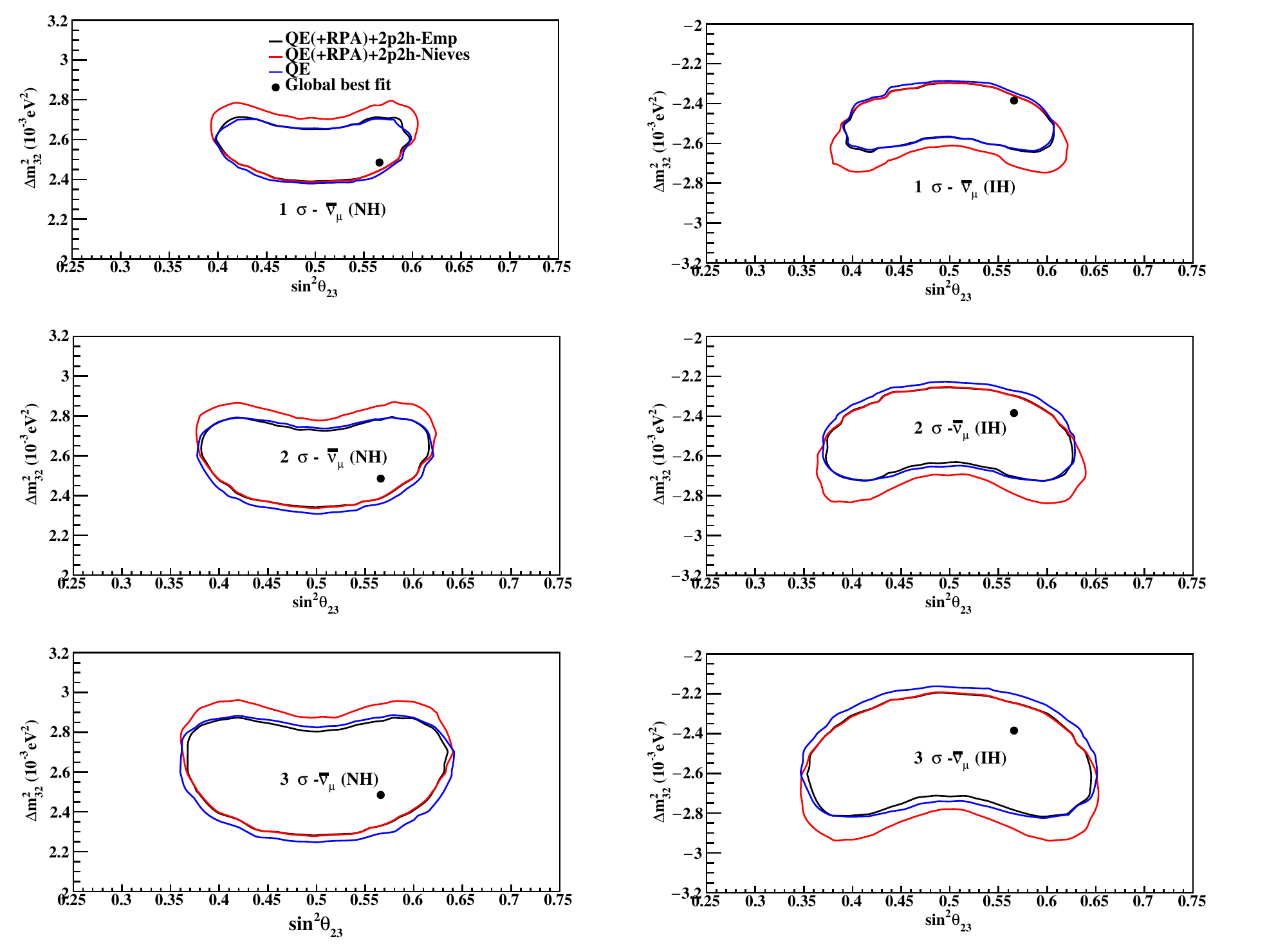}
 \caption{Comparison of 1$\sigma$, 2$\sigma$ and 3$\sigma$ contours for $\Delta m^{2}_{32}$ vs $\sin^{2}\theta_{23}$ for antineutrino without detector response.} 
\end{figure}

\section{Results and Discussion}
\label{sec:4}
In this section, we present the results in Figs. (6-11) on sensitivity analysis of the $\nu_{\mu}$ and $\bar\nu_{\mu}$ disappearance channel at NO$\nu$A, and also a discussion on them. The extrapolated neutrino and antineutrino FD events prediction are computed in a three flavor neutrino oscillation scenario where the $\nu$ and $\bar\nu$ parameters - the atmospheric mass splitting $\Delta m^{2}_{32}$ and the mixing angle $\theta_{23}$ are allowed to vary independently. We compare sensitivity results separately for 1$\sigma$, 2$\sigma$, and 3$\sigma$ contours for QE without RPA and 2p2h, and QE(+RPA)+2p2h interactions for NH (left panel) and IH (right panel) for neutrino in Fig. 6, and antineutrino in Fig. 7, without detector effect to observe the differences due to MN interactions. QE without RPA and 2p2h with/without detector effect for neutrino and antineutrino are shown in Figs. 8 and 9. In all these Figs. (6-11), left panel is for NH, while right panel is for IH. Figs. 10 and 11 are shown for comparison of Empirical and Nieves et al. model for 2p2h in QE(+RPA)+2p2h interactions with no detector effect (black line), detector effect with efficiency 31.2\% ($\nu_{\mu}$) or 33.9\% ($\bar\nu_{\mu}$) (red line), 50\% (blue line) and 80\% (magenta line).

\begin{figure}[H] 
\centering\includegraphics[width=17cm, height=12cm]{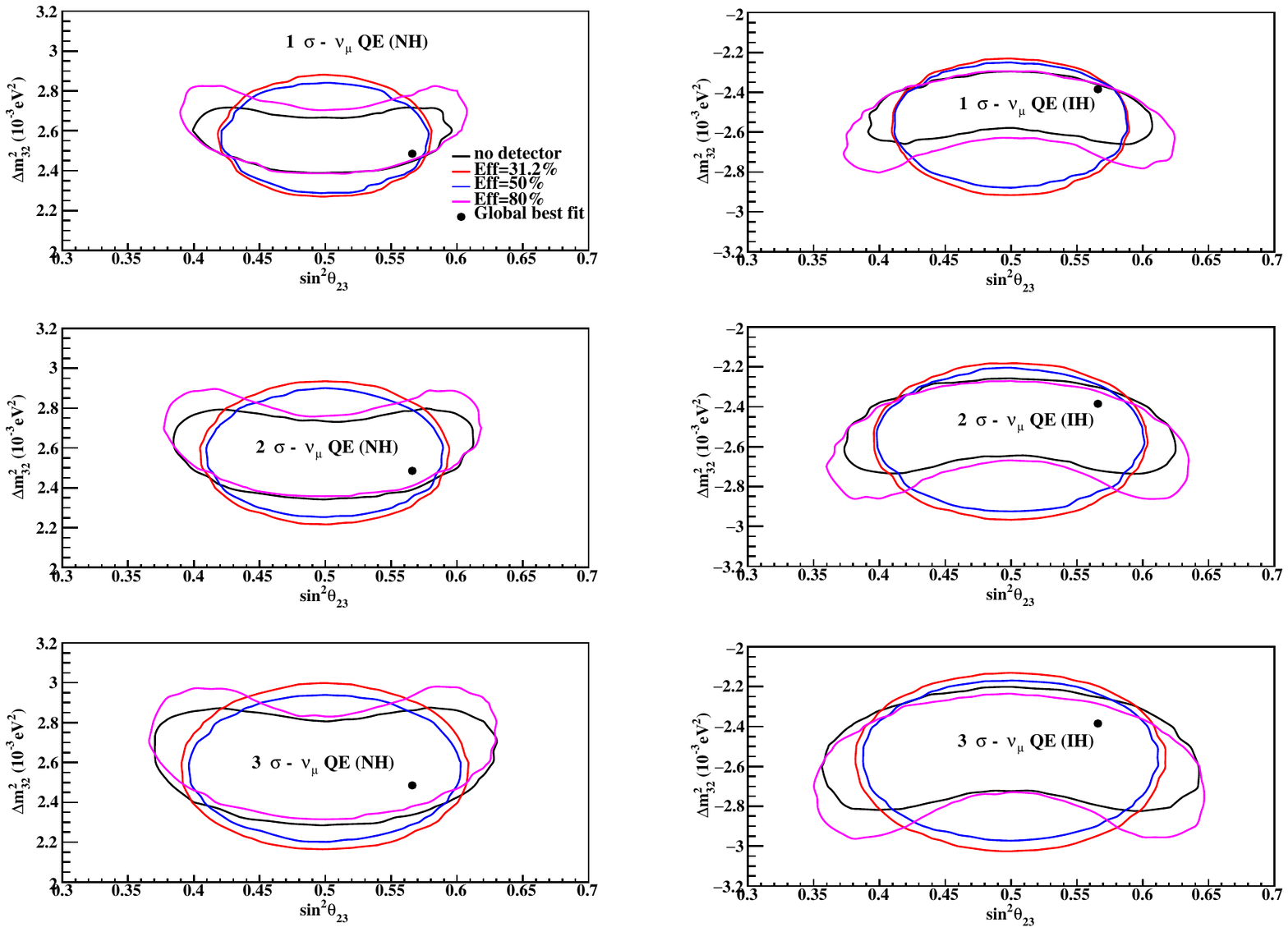}
\caption{Comparison of  1 $\sigma$, 2$\sigma$ and 3$\sigma$ contours in $\Delta m^{2}_{32}$ vs $\sin^{2}\theta_{23}$  for QE without RPA and 2p2h for NH (left panel) and IH (right panel) for neutrino mode with/without detector effect. Here black solid line represents the no detector effect, red solid line detector effect with 31.2\% efficiency, blue solid line 50\% efficiency and magenta solid line 80\% efficiency.}
\end{figure}

\begin{enumerate}

\item In Fig. 6 and Fig. 7, we have shown the comparison of 1$\sigma$, 2$\sigma$ and 3$\sigma$ contours separately in $\Delta m^{2}_{32}$ vs $\sin^{2}\theta_{23}$ plane for the interactions QE(+RPA)+2p2h (Empirical, black line), QE(+RPA)+2p2h (Nieves et al., red line), QE without RPA and 2p2h (blue line) for NH (left panel) and IH (right panel) for both neutrino and antineutrino without detector response. We notice that contours due to QE(+RPA)+2p2h (Empirical, black solid line) show slightly less uncertainty (less area inside contour) than contours due to QE(+RPA)+2p2h (Nieves et al., red solid line) and QE (blue line) in both neutrino and antineutrino cases. This can be attributed to the shift of the peak of the first oscillation maximum in the Empirical model case towards little higher energy as compared to the Nieves et al.  model (please see Fig. 4 and Fig. 5). There is a clear difference between the three interactions in 1$\sigma$, 2$\sigma$ and 3$\sigma$ regions in case of neutrino, but for antineutrino, contours of  QE(+RPA)+2p2h (Empirical) and QE mostly overlap in 1$\sigma$, 2$\sigma$ regions while there is a slight difference in 3$\sigma$. This implies that these MN effects cause lesser change in antineutrino case than in neutrino. It is known that correct identification of neutrino energy is a prerequisite  for precise measurement of neutrino oscillation parameters. Since amplitude of the oscillation provides measurement of the mixing angle, while its phase is translated to the value of neutrino mass squared difference, hence it is clear that the changes in probability due to MN interactions (as seen in Figs. 4 and 5) causes a corresponding shift of the oscillation parameters (as seen in Figs. 6 and 7). This discussion is also applicable to all the cases below.

\item From Figs. 8 and 9, it is observed that the sensitivity contours (for pure QE process with no RPA and 2p2h effects) with 80\% detector efficiency are the closest to those with no detector effect (no detector effect  indeed corresponds to 100\% detector efficiency and no inclusion of the resolution function for detector) compared to 50\%, 31.2\% (neutrino) and 33.9 \% (antineutrino) efficiency. This is as expected - more is the efficiency of the detector, more precise would be its measurements (less uncertainty) of oscillation parameters, and hence lesser is the allowed region of parameter space. 

\begin{figure}[H] 
 \centering\includegraphics[width=17cm, height=12cm]{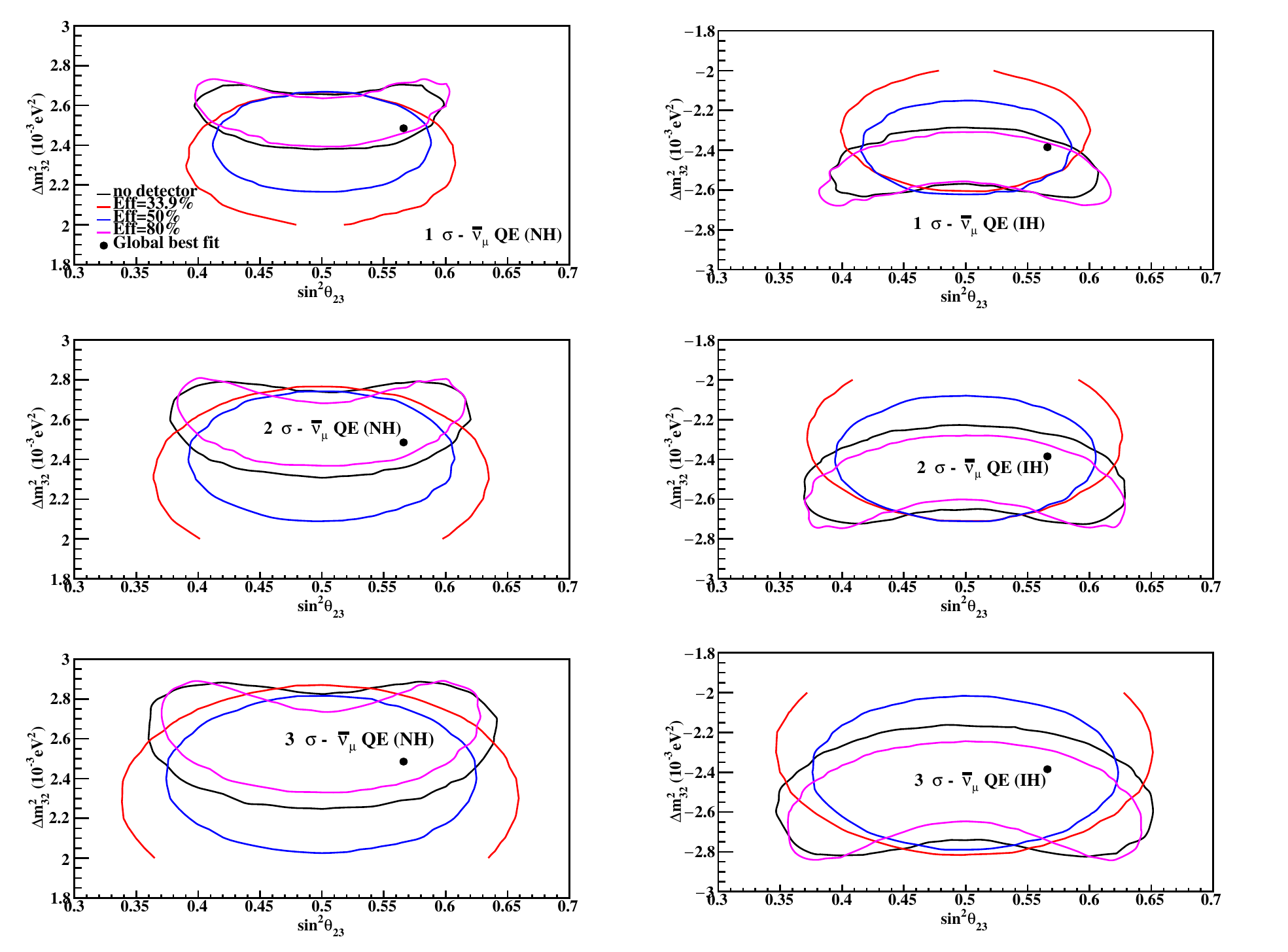}
 \caption{Comparison of  1 $\sigma$, 2$\sigma$ and 3$\sigma$ contours in $\Delta m^{2}_{32}$ vs $\sin^{2}\theta_{23}$  for QE without RPA and 2p2h for NH (left panel) and IH (right panel) for antineutrino mode with/without detector effect. Here black solid line represents the no detector effect, red solid line detector effect with 33.9\% efficiency, blue solid line 50\% efficiency and magenta solid line 80\% efficiency.}
\end{figure}

\item Results in Figs. 10 and 11, show that out of  the two models of 2p2h, QE(+RPA)+2p2h with Empirical model shows a better result than QE(+RPA)+2p2h with Nieves et al. model for both neutrino and antineutrino case and for both the mass hierarchies with and without detector effect. Also, as observed in previous point, sensitivity contours with 80\% detector efficiency are the closest to those with no detector effect.

\item Also, it can be seen that the areas enclosed by the contours without detector effect, before and after including MN effects are more in antineutrino than neutrino - which again can be attributed to less flux and cross-section for antineutrino. And, there is more separation between contours of with and without MN interactions (reflection of Fig. 4 and 5), for neutrino case, as compared to that of antineutrino. So, we can say that MN interactions affect neutrino scattering more than that of antineutrino \cite{Martini:2010ex}.

\end{enumerate}

\section{Summary and Conclusion}
\label{sec:5}

To summarise, in this work, we studied the effect of MN interactions on the sensitivity of measurement of neutrino oscillation parameters atmospheric mixing angle $\theta_{23}$ and $\Delta m^{2}_{32}$, at NO$\nu$A experiment (Carbon target), using disappearance channel, and also computed the impact of realistic detector effects on extraction of oscillation parameters. In particular, we considered RPA and 2p2h interactions. We used event generator GENIE for the purpose, along with extrapolation technique for computing FD events spectrum and Feldman's technique for the $\chi^{2}$ analysis. For 2p2h modelling, we considered both the Empirical and Valencia (Nieves et. al.) models. Migration matrices were generated and used for both ND and FD. We did the analysis with/without detector effects, and for both the hierarchies (NH and IH). The global best fit central values from \cite{deSalas:2020pgw} were used as a true set for the $\chi^{2}$ analysis, while their $3\sigma$ range was used for simulated experiments (test values). We used kinematical method of energy reconstruction, and 1 million MC events were generated for simulation at ND and FD. MN interactions affect the neutrino-nucleus scattering cross section and events spectrum - they change the amplitude and position of peaks of the FD events spectrum. We also observed that the 2p2h Empirical model shows a better result in sensitivity analysis than the 2p2h Nieves et al. model. Both for neutrino and antineutrino, there is clear distinction among the three curves - pure CCQE, QE(+RPA)+2p2h (Empirical), QE(+RPA)+2p2h (Nieves et al.) in medium energy ranges, say 0.5 $\leqslant E_{\nu} \leqslant$ 3 GeV and 0.5 $\leqslant E_{\nu} \leqslant$4.5 GeV respectively. Cut-offs towards the high energy tail may also have a role to play. It was observed that the MN effects change the cross-section for neutrino more, than for antineutrino, and this feature was noted earlier also \cite{Martini:2010ex}.

\begin{figure}[H] 
 \centering\includegraphics[width=17cm, height=12cm]{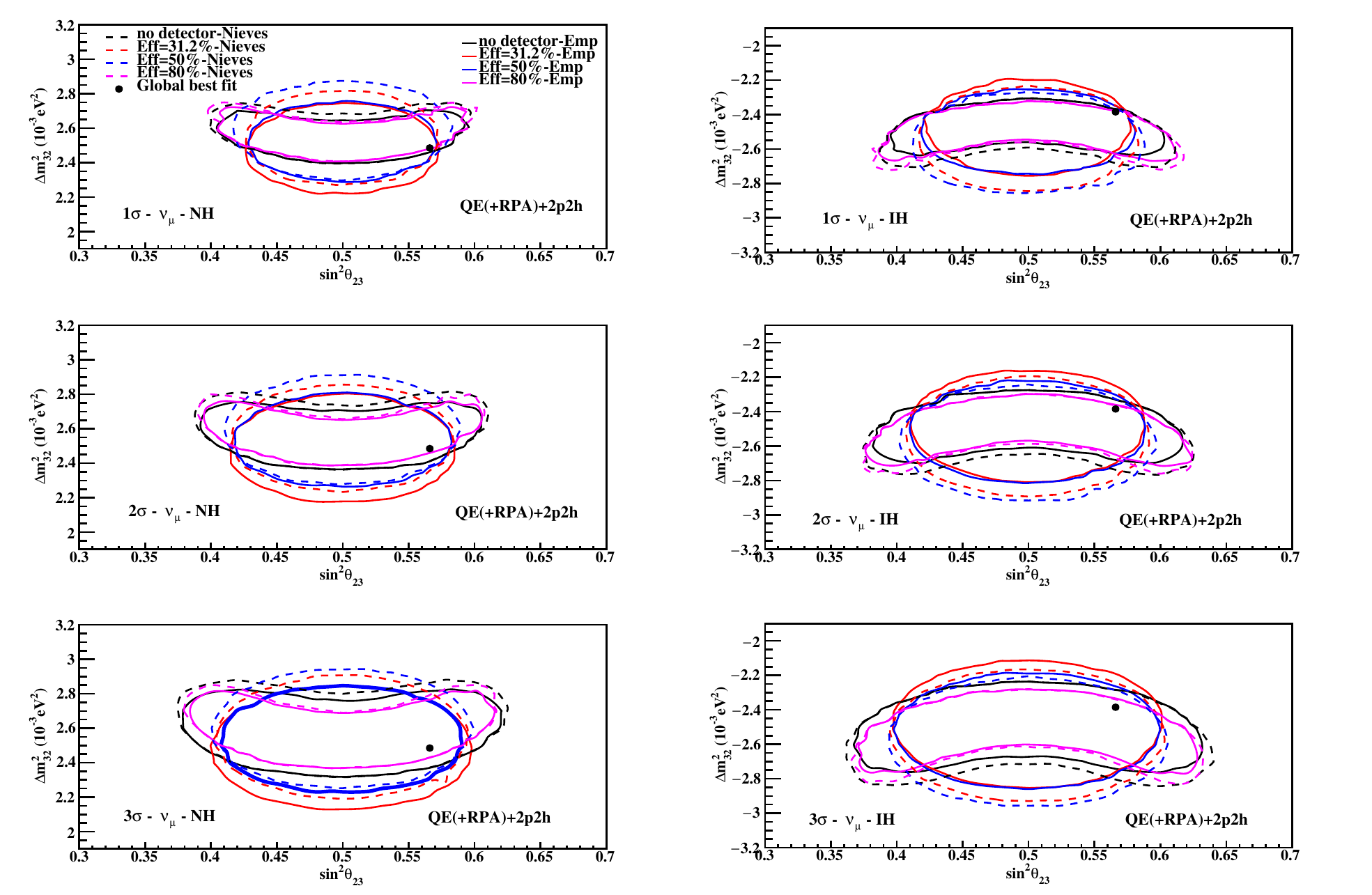}
 \caption{Comparison of 1 $\sigma$, 2$\sigma$ and 3$\sigma$ contours in $\Delta m^{2}_{32}$ vs $\sin^{2}\theta_{23}$  for QE(+RPA)+2p2h(Empirical, solid line) and QE(+RPA)+2p2h(Nieves et al., dash line) for NH (left panel) and IH (right panel) for neutrino mode with/without detector effect. Here black solid line represents the no detector effect, red solid line detector effect with 31.2\% efficiency, blue solid line 50\% efficiency and magenta solid line 80\% efficiency.}
\end{figure}

It is known that correct reconstruction of energy of neutrino is very important to extract the pinpointed value of neutrino oscillation parameters. And hence this shift in oscillation probability due to MN interactions  affects the sensitivity in the measurement of oscillation parameters. The main results of this work were shown in Figs 6-11 as sensitivity contours in ($\theta_{23}-\Delta m^{2}_{32}$) plane, which clearly showed that MN interactions do affect the sensitivity analysis, which is a manifestation of shift in the amplitude and phase in oscillation probability (reflected in measurements in the experiments). The areas inside the sensitivity contours before and after including MN effects are more in antineutrino than neutrino - which again can be attributed to less flux and cross-section for antineutrino (and hence more uncertainty due to less data). Results were presented for both hierarchies, and respective changes are seen in both of them. The sensitivity contours with 80\% efficiency were found to be the closest to those with no detector effect (which corresponds to 100\% detector efficiency and no resolution). The lower efficiency of the detector implies more uncertainty in sensitivity studies and this stresses the need for improvement in detector efficiencies in future.  Similar effects are also expected to appear in the $\nu_{\mu}\rightarrow\nu_{e}$ appearance channels as well. These can also impact the sensitivity to CP violation at the future long-baseline experiments as the CP violation analysis depends on energy of the signal and difference between neutrino and antineutrino rates. We also aim to tune the Empirical model in the way as close as possible to the one followed by NO$\nu$A collaboration in the oscillation analyses. To conclude, we still do not fully understand the uncertainties in the modelling of MN interactions and their implementations in the Monte Carlos in neutrino oscillation experiments, and hence the studies done in this work are important.  Though future measurements of neutrino-nucleus cross sections will tell us, which nuclear interaction model is better, this study however surely indicates that MN interactions have a significant effect on sensitivity measurements, and hence command a careful inclusion of them in such analyses. 

\begin{figure}[H] 
 \centering\includegraphics[width=17cm, height=12cm]{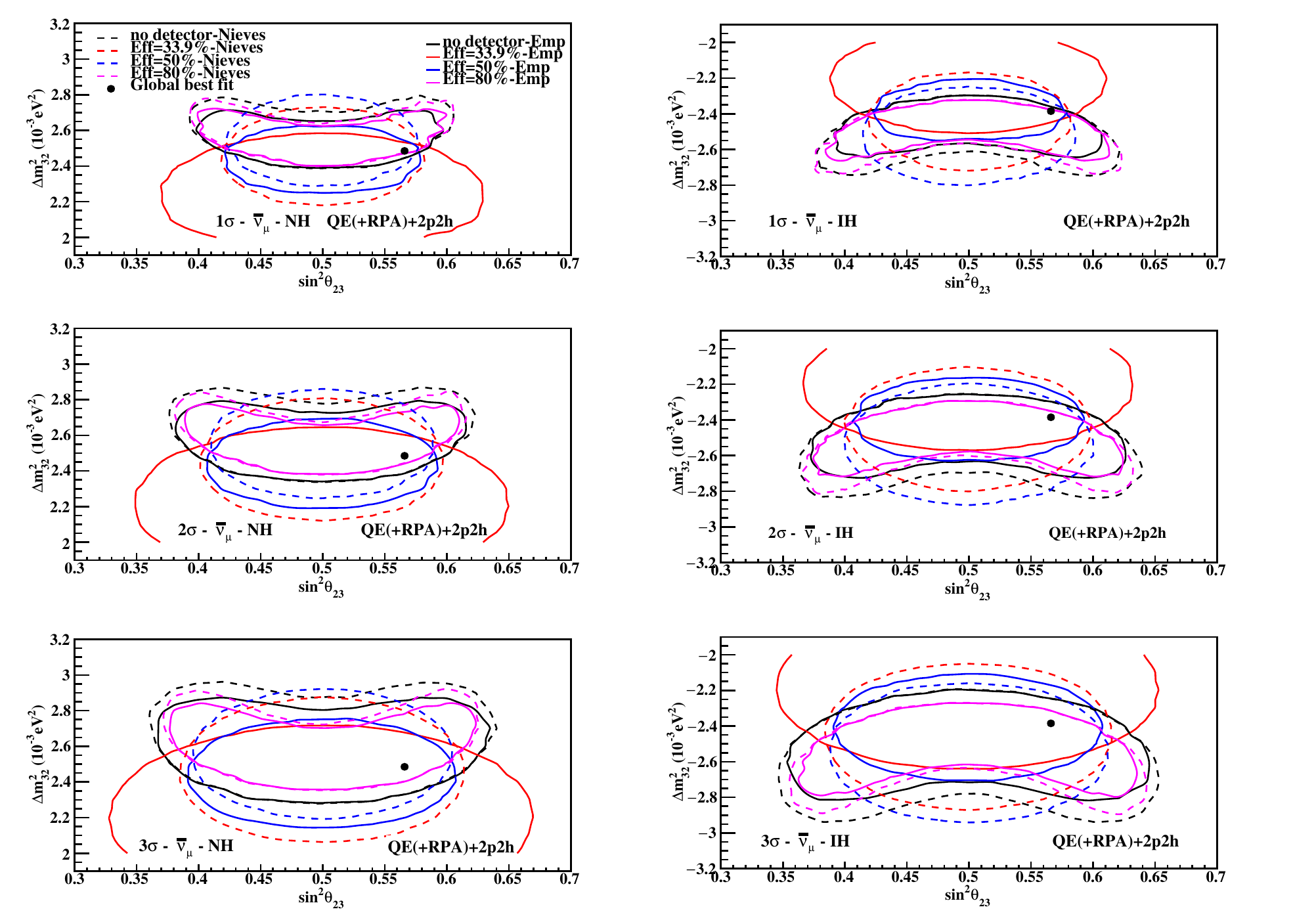}
 \caption{Comparison of 1 $\sigma$, 2$\sigma$ and 3$\sigma$ contours in $\Delta m^{2}_{32}$ vs $\sin^{2}\theta_{23}$  for QE(+RPA)+2p2h(Empirical, solid line) and QE(+RPA)+2p2h(Nieves et al., dash line) for NH (left panel) and IH (right panel) for antineutrino mode with/without detector effect. Here black solid line represents the no detector effect, red solid line detector effect with 31.2\% efficiency, blue solid line 50\% efficiency and magenta solid line 80\% efficiency.}
\end{figure}

\section{Acknowledgments} 
\label{sec:6}

We sincerely thank the honourable referee, for making very fruitful and constructive comments, that has helped us to improve the work. KB thanks DST-SERB, Govt. of India, for support through her major project EMR/2014/000296, during which a part of this work was done. KB thanks Prof. U. Mosel and Dr. J. Paley for clarification on units/normalisation of neutrino flux file, over email. She also would like to acknowledge some support from Gauhati University in her laboratory. KB, NS and JS sincerely thank Prof. Raj Gandhi of HRI, Prayagraj, India for inspiring discussions during the initial stages of the work. KB and PD would like to acknowledge support from RUSA and FIST grants of Govt. of India in upgrading the computer laboratory of the Physics department of GU where part of this work was done.

\end{document}